\documentclass[10pt]{article}
\usepackage[dvips]{graphicx,psfrag}
\usepackage{amssymb}
\usepackage{color} 
\usepackage{setspace}
\input epsf
\definecolor{Blue}{rgb}{0.3,0.3,0.9}
\definecolor{Red}{rgb}{1,0,0}
\definecolor{Green}{rgb}{0,1,0}
\newcommand{\be}{\begin{equation}}
\newcommand{\ee}{\end{equation}}
\newcommand{\bea}{\begin{eqnarray}}
\newcommand{\eea}{\end{eqnarray}}

\begin{document}
\begin{titlepage}

\begin{center}
{\Large\bf A WKB approach to scalar fields dynamics in curved space-time}

\bigskip \bigskip \medskip
{\large J. Grain \& A. Barrau\footnote{corresponding author, email :
Aurelien.Barrau@cern.ch}
}\\[.5cm]
{\it Laboratory for Subatomic Physics and Cosmology, Joseph Fourier University,
CNRS-IN2P3, 53, avenue de Martyrs, 38026 Grenoble cedex, France }\\[1cm]

{\bf Abstract}
\end{center}
Quantum fields exhibit non-trivial behaviours in curved space-times, especially
around black holes or when a cosmological constant is added to the
field equations. A new scheme, based on the Wentzel-Kramers-Brillouin (WKB) 
approximation is presented. The main advantage of this method is to allow for a
better physical understanding of previously known results and to give good
orders of magnitude in situations where no other approaches are currently
developed. Greybody factors for evaporating black holes are rederived in this
framework and the energy levels of scalar fields in the Anti-de Sitter (AdS) spacetime
are accurately obtained. Stationary solutions in the Schwarzschild-Anti-de Sitter (SAdS)
background are investigated.
Some improvements and the basics of a line of thought for more complex situations are
suggested.\\


\end{titlepage}
%\tableofcontents
\section{Introduction}

The study of some of the most intriguing phenomena in our Universe, such as the
evaporation of microscopic black
holes or the parametric amplification of primordial quantum fluctuations, 
is performed within the framework of quantum field theory in curved space-time. Although
well-established and reliable, as it does not require any quantization of the gravitational 
field itself, this theoretical framework leads to highly non-trivial partial differential equations 
whose resolution is, in most cases, based on numerical investigations. The Wentzel-Kramers-Brillouin 
(WKB) approximation and its extension to semi-classical methods (introduced by Kramers
\cite{kramers}, Young \& Uhlenbeck \cite{young} and readdressed by Langer \cite{langer})
can be extremely usefull to describe those physical systems. 

The WKB method is used to approximately solve the 
following equation~:
\begin{equation}
	\left(\frac{d^2}{d\xi^2}+p^2(\xi)\right)\psi(\xi)=0
	\label{equ}
\end{equation}
where the real function $p^2$ can be either positive or negative valued.
Nevertheless, WKB wave functions $\tilde{\psi}$ are exact solutions of another
differential equation~:
\begin{equation}
\left(\frac{d^2}{d\xi^2}+p^2(\xi)\right)\tilde{\psi}(\xi)=\left(\frac{3}{4p^2}\left(\frac{dp}{d\xi}\right)^2-\frac{1}{2p}\frac{d^2p}{d\xi^2}\right)\tilde{\psi}(\xi),
	\label{wkb-equ}
\end{equation}
which restricts the validity of the approximation to regions where
\begin{equation}
	\left|p^2\right|\gg\left|\frac{3}{4p^2}\left(\frac{dp}{d\xi}\right)^2-\frac{1}{2p}\frac{d^2p}{d\xi^2}\right|.
	\label{wkb-val}
\end{equation}
It means that if the momentum is a linear 
function of $\xi$, 
the WKB wave function becomes an exact solution of Eq.~(\ref{equ}). The WKB approximation can 
therefore provide very accurate 
results if the considered momentum squared is,
depending on the potential, dominated by the constant, linear and quadratic terms in its Taylor 
expansion.

Recently, this approximation has been used to derive  many 
important results about quantum fields propagating in a curved background.
The Hawking evaporation process \cite{parikh}, as well as the inflationary 
cosmological perturbations \cite{martin,casadio}, have been investigated within this scheme. 
In Ref.~\cite{parikh}, the 
Hawking temperature for a Schwarzschild black hole was rederived
whereas, in Ref.~\cite{martin}, the spectral index and the amplitude of the inflationary 
perturbations were recovered nearly exactly. 
Finally, the WKB method has also been extensively used in order to determine the quasi-normal 
frequencies of black holes (see Ref.~\cite{nollert} and 
Ref.~\cite{qnm}). The WKB tunnel probability used in the computation of the
quasi-normal frequencies has also been applied to estimate the greybody factors for gravitons 
emitted by a $D-$dimensional Schwarzschild black holes \cite{cornell}.

In this article, we extend the idea of the WKB and semi-classical methods
to the 
propagation of massless, minimally coupled, scalar fields in a static and spherically symmetric 
curved background, called Schwarzschild-like hereafter. Section 2 is devoted to the
derivation of the radial part of the Klein-Gordon (KG) equation and to its rewriting in a 
Schr\"odinger-like form, such as Eq.~(\ref{equ}). It is shown that this new equation is the 
quantum version of a 
slightly-modified relativistic Hamilton-Jacobi equation. The 
semi-classical scheme is applied to solve the problem and to derive the
quantization rule and WKB tunnel probability. In Section 3, the discrete energy 
spectrum for massless scalar fields embedded in an Anti-de Sitter (AdS) Universe is derived 
and compared with the exact spectrum \cite{ads-level,private}. 
In Section 4, the problem of the determination of the greybody factors, {\it i.e.} the 
coupling between quantum fields and black holes, is investigated. 
The emission/absorption cross-sections are thus determined for scalar fields evolving 
in a $4-$dimensional Schwarzschild background. The resulting WKB cross-sections are also compared
with the exact ones (and to those obtained in Ref.~\cite{cornell}). 
In Section 5, the propagation of scalar fields in a Schwarzschild-Anti-de Sitter (SAdS) space-time 
is addressed~: it is shown that resonances, whose energies and bandwidths are obtained, arise in the state density. Finally, in Section 6, we draw up some conclusions and comments about the usefulness and 
reliability of the semi-classical approximation and underline some perspectives to improve 
this method. Some general arguments about its significance to study quantum fields in curved 
space-time are briefly given.

\section{Scalar fields in Schwarzschild-like space-time}

	\subsection{Equation of motion}
	\label{kg-part}
	
A static and spherically symmetric space-time is described by the general
Schwarzschild-like metric, totally determined by one function $h(r)$~:

\begin{equation}
	ds^2=h(r)dt^2-\frac{dr^2}{h(r)}-r^2d\Omega^2
\end{equation}
where $d\Omega$ is the solid-angle. 
The propagation of massless and minimally coupled scalar particles in this 
background is given by the solution of the KG equation $\Phi^{;\mu}_{~;\mu}=0$. 
The field can be 
decomposed into eigenmodes of normal frequency $\omega$ and angular momentum
$\ell$, 
$\Phi(t,r,\Omega)=e^{-i\omega{t}}Y^{m}_{\ell}(\Omega)R(r)$, the 
dynamics being totally given by the radial function $R(r)$ which
satisfies the radial part of the KG equation~:
\begin{equation}
	\frac{d}{dr}\left(hr^2\frac{dR}{dr}\right)+\left(\frac{\omega^2r^2}{h}-\ell(\ell+1)\right)R=0
	\label{kg}
\end{equation}
where $\ell(\ell+1)$ corresponds to the eigenvalues of the angular part of the KG 
equation. 

	The resolution of this equation strongly depends on the shape of the 
metric function $h$ and it is usually not possible to obtain exact solutions, even in 
a simple background such as the Schwarzschild space-time. However, Eq. (\ref{kg}) 
can be modified by the use of the
tortoise coordinate $r^\star$, defined as $dr^\star=dr/h(r)$, and a new radial 
function $U(r)=rR(r)$. It becomes of the Schr\"odinger type~:
\begin{equation}
	-\frac{d^2U}{d{r^\star}^2}+\underbrace{h(r)\left(\frac{\ell(\ell+1)}{r^2}+\frac{1}{r}\frac{dh}{dr}\right)}_{{\mathcal{M}}(r^\star)}U=\omega^2U,
	\label{schro}
\end{equation}
which means that the square of the momentum operator $i\partial/\partial{r^\star}$ 
is equal to $\omega^2-{\mathcal{M}}(r^\star)$. We interpret Eq.~(\ref{schro}) as the 
operational version of the classical Hamilton-Jacobi equation 
${\mathcal{H}}^2=p^2+V^2(r^\star)$ where the ${\mathcal{M}}$ function is identified 
to the square of a potential-like contribution. Except by the author of Ref. 
\cite{chandra}, the ${\mathcal{M}}$ function is usually associated with $V$
and not with $V^2$ (see, {\it e.g.}, Ref.~\cite{qnm,frolov}). Although this does 
not change the mathematical structure and the solutions, we prefer to identify 
${\mathcal{M}}$ 
function with the square of the potential in order to obtain a homogeneous classical 
Hamilton-Jacobi equation in a physically-consistant way. Eq. (\ref{schro}) 
can be regarded as the quantum version of the classical equation of motion for a test 
particle in a Schwarzschild background given by~:
\begin{equation}
	\left(\frac{dr}{d\tau}\right)^2+h(r)\left(1+\frac{L^2}{r^2}\right)=E^2
\end{equation}
where the second term of the left hand side of the above equation is seen as 
the square of the potential. Consequently, the propagation of scalar fields in a 
Schwarzschild-like background is based on a classical problem described by the action 
$S$ solution of
\begin{equation}
	\left(\frac{\partial{S}}{\partial{r^\star}}\right)^2+V^2(r^\star)=\left(\frac{\partial{S}}{\partial{t}}\right)^2.
	\label{ham-jac}
\end{equation}
It should be noticed that the above 
equation slightly differs from the relativistic Hamilton-Jacobi equation for massless
particles evolving in a spatial potential. Particles propagating in a spatial 
potential are described by the Lagrangian
$
	{\mathcal{L}}=-m\sqrt{1-v^2}-V(r^\star)
$
whereas the considered problem can be seen as a free particle with a mass 
which depends upon the position
$
	{\mathcal{L}}=-V(r^\star)\sqrt{1-v^2}.
$
Because Eq. (\ref{schro}) has exactly the same mathematical form as Eq. (\ref{equ}),
the semi-classical approximation can be easily applied to the propagation of 
massless scalar fields in a Schwarzschild-like background, keeping in mind that the 
underlying classical problem is described by the above-mentioned Lagrangian.

	\subsection{Semi-classical approach}
	\label{wkb-part}

	 The quantum version of the Hamilton-Jacobi relation is given by 
the following KG-like equation~:
\begin{equation}
	\left(-\hbar^2\frac{\partial^2}{{\partial{\xi}}^2}+V^2(\xi)\right)\psi(\xi,t)=-\hbar^2\frac{\partial^2\psi}{{\partial{t}}^2}(\xi,t).
	\label{schro-equ-tps}
\end{equation}
In a Schwarzschild-like background, this equation corresponds to the time and 
radial part of the KG equation, the angular part being separated using 
spherical harmonics, with $\xi$ standing for the tortoise radial coordinate and 
$t$ for the Schwarzschild time. The $U$ radial function of eq. \ref{schro}
corresponds to the spatial part of the $\psi$ wavefunction in the above equation.
For an initial event $\mathcal{I}=(t_i,\xi_i)$ 
and a final one 
$\mathcal{F}=(t_f,\xi_f)$, the ansatz corresponds to the sum over all the classical 
paths given by
\begin{equation}
	\psi({\mathcal{F}},{\mathcal{I}})=\displaystyle\sum_nF_n({\mathcal{F}},{\mathcal{I}})e^{i\frac{S_n({\mathcal{F}},{\mathcal{I}})}{\hbar}},
\end{equation}
where $S_n({\mathcal{F}},{\mathcal{I}})$ is the classical action of the $n^{th} $
path between $\mathcal{I}$ and $\mathcal{F}$. Performing the power expansion  up to the linear 
order leads to~:
\begin{eqnarray}
	\left\{\begin{array}{l}
		i\hbar\left(2\frac{\partial{S}}{\partial{\xi}_f}\frac{\partial{F}}{\partial{\xi}_f}+\frac{\partial^2S}{{\partial{\xi}_f}^2}F\right) \\
		+\left(\left(\frac{\partial{S}}{\partial{\xi}_f}\right)^2+V^2\right)F
	       \end{array}\right\} & = & \left\{\begin{array}{l}
						i\hbar\left(2\frac{\partial{S}}{\partial{t}_f}\frac{\partial{F}}{\partial{t}_f}+\frac{\partial^2S}{{\partial{t}_f}^2}F\right) \\
						+\left(\frac{\partial{S}}{\partial{t}_f}\right)^2F
	       \end{array}\right\}
\end{eqnarray}
where the $n$ indices have been dropped for simplicity. The resolution is not 
directly performed and an alternative set of equations should be used.
The first order relation leads to~:
\begin{equation}
	\frac{\partial}{\partial{\xi_f}}\left(\left|F\right|^2\frac{\partial{S}}{\partial{\xi_f}}\right)-\frac{\partial}{\partial{t_f}}\left(\left|F\right|^2\frac{\partial{S}}{\partial{t_f}}\right)=0.
	\label{continuity}
\end{equation}
Considering Eq. (\ref{continuity}) as an equation of continuity, we can define the 
quadri-current at the semi-classical order as  $\left|F\right|^2 p_\mu$, so that~:
\begin{equation}
	-\left|F\right|^2p_\mu=i\left(\psi\partial_\mu\psi^\dag-\psi^\dag\partial_\mu\psi\right).
\end{equation} 
%A set of solution for the $\left|F\right|^2$ function is given by 
%\begin{equation}
%	\left|F\right|^2=.
%\end{equation}
%It can be checked that applying the $$ operator to classical Hamilton-Jacobi equation leads to Eq. (\ref{continuity}) where the above solution isntead of the $\left|F\right|^2$ function.
%as the $F$ function gives the semi-classical probability amplitude, the $\left|F\right|^2\frac{\partial{S}}{\partial{x_f}}$ and the $\left|F\right|^2\frac{\partial{S}}{\partial{t_f}}$ functions are respectively interpreted as a probability current and a probability density. which leads to a new constrain determining the phase of the $F$ function 
%\begin{equation}
% 	\ln{\left(\frac{F}{F^\dag}\right)}=-iS.
%	\label{phase}
%\end{equation}
%fixing the phase $\phi=-S/2$.

For a stationary problem, {\it{i.e}} 
${\partial^2\psi}/{{\partial{t}}^2}=-\omega^2\psi$, it is more convenient to consider
a path with fixed energy and the Green function of the propagator is preferred. 
Using the stationary phase approximation, it is 
shown that, as in non-relativistic quantum
mechanics, the Green function takes the form
\begin{equation}
G(\xi_f,\xi_i,\omega)=\displaystyle\sum_n\sqrt{\frac{p_n(\xi_i)}{p_n(\xi_f)}}e^{i\int_{\xi_i}^{\xi_f}p_n(\xi)d\xi},
\end{equation}
where $p_n=\frac{\partial{S_n}}{\partial{\xi}}$, corresponds to the classical 
momentum $\sqrt{\omega^2-V^2(\xi_f)}$ at the energy $\omega$ and does not depend upon 
the considered path. Althought most non-relativistic quantum results can be
easily adapted, we have carefully checked, thanks to an explicit calculation of the density
of states, that the Bohr-Sommerfeld quantization rule in a well-potential is recovered~:
\begin{equation}
	2\displaystyle\int_{\xi_-(\omega_n)}^{\xi_+(\omega_n)}p_n(\xi,\omega_n)d\xi=\left(2n+1\right)\pi,
	\label{bohr}
\end{equation}
where $n$ is an integer and $\xi_\pm$ the two turning points. Following the techniques of 
Ref. \cite{messiah} (the classical action is here
allowed to become complex) and using the matching procedure 
at the turning point described in appendix \ref{app-c}, the transmission coefficient at the
WKB order is found to be
\begin{equation}
	\left|A\right|^2=\exp{\left(-2\int^{\xi_+}_{\xi_-}p'(\xi,\omega)d\xi\right)}
\end{equation}
where the interval $[{\xi_-},{\xi_+}]$ corresponds to the region where  energy
of the
particle is lower than the potential.

\section{Energy spectrum in AdS Universe}
\label{ads}

\subsection{Exact results}
	
	It was shown in \cite{ads-level} that massless 
scalar fields, minimally coupled to gravity, propagating in an AdS space-time 
exhibit a discrete energy spectrum. We will briefly rederive those exact results using 
another --less rigourous-- approach to compare them with semi-classical computations.
For scalar particles 
propagating in a pure AdS space-time, the following Schwarzschild-like metric 
has to be considered :
\begin{equation}
	ds^2=\left(1+\frac{r^2}{R^2}\right)dt^2-\frac{dr^2}{\left(1+\frac{r^2}{R^2}\right)}-r^2d\Omega^2
\end{equation}
where $R=\left(-\Lambda/3\right)^{1/2}$ is the curvature radius of the AdS 
space-time. The KG equation is solved \cite{private} using a 
more convenient coordinate system \cite{hawk_book,isham}~:
\begin{equation}
	\begin{array}{l}
		\tau=\frac{t}{R}~;~
		\rho=2\arctan\left(\frac{r}{R}+\sqrt{1-\frac{r^2}{R^2}}\right)-\frac{\pi}{2}~;~
		ds^2=\frac{R^2}{\cos^2(\rho)}\left(d\tau^2-d\rho^2-\sin^2{(\rho)}d\Omega^2\right).
	\end{array}
	\label{tau-t}
\end{equation}
With the ansatz
$\Phi(\tau,\rho,\Omega)=e^{-i\nu{\tau}}Y^{m}_{\ell}(\Omega)R(\rho)$, the following
 radial equation is obtained~:
\begin{equation}
\frac{\cos^2{(\rho)}}{\sin^2{(\rho)}}\frac{d}{d\rho}\left(\frac{\cos^2{(\rho)}}{\sin^2{(\rho)}}\frac{dR}{d\rho}\right)+\left(\nu^2-\frac{\ell(\ell+1)}{\sin^2{(\rho)}}\right)R=0
	\label{eq-ads}
\end{equation}
which leads --see appendix \ref{app-a}-- to~:
\begin{equation}
R(\rho)=\sin^\ell{(\rho)}\cos^3{(\rho)}P_{(\nu-\ell-3)/2}^{(\ell+1/2,3/2)}\left(\cos{(2\rho)}\right).
\end{equation}

This solution, based on Jacobi polynomial, leads to the energy spectrum 
\begin{equation}
	\left(\omega_{n,\ell}\right)_{(n,\ell)\in{\mathbb{N}}^2}=\left(\frac{2n+\ell+3}{R}\right)_{(n,\ell)\in{\mathbb{N}}^2}
	\label{ener-spec}
\end{equation}
where a zero point energy clearly appears. The curvature of 
space-time, $1/R$, provides a natural unit for scalar particles energy and this quantization can be viewed as periodic conditions since the geodesics are 
closed in the AdS Universe. 

	\subsection{The Bohr-Sommerfeld quantization rule}

\begin{figure} 
	\begin{center}
		\includegraphics[scale=0.6]{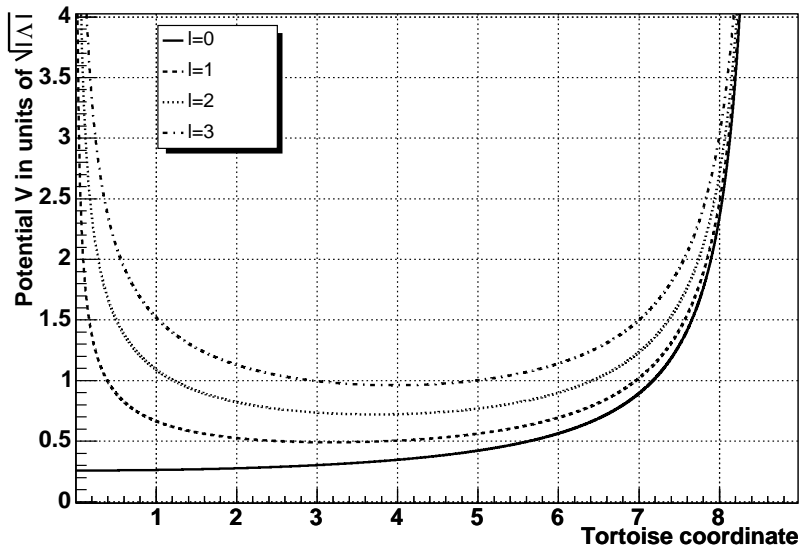}
		\caption{Potential seen by a scalar field in AdS space-time in units of $\sqrt{\left|\Lambda\right|}$.}
		\label{pot-ads-pure}
	\end{center}
\end{figure}

The above results are a direct consequence of the quantum nature of the fields 
combined with the intrinsic curvature of the AdS Universe. Rewriting the radial 
part of the KG equation in a Schr\"odinger-like form, as mentioned in section 
\ref{kg-part}, it is clear that the potential seen by a scalar particle is a
well-potential for 
all the values of the orbital momentum but zero. This is illustrated in Fig. 
\ref{pot-ads-pure} and given by
\begin{equation}
	V^2_\ell(r^\star)=\frac{1}{R^2\cos^2{\left(r^\star/R\right)}}\left(\frac{\ell(\ell+1)}{\tan^2{\left(r^\star/R\right)}}+2\right)
\end{equation}
where the tortoise coordinate $r^\star=R\arctan{\left(r/R\right)}$ lies in the 
range $[0,\pi{R}/2[$. As a consequence, scalar fields clearly have to 
exhibit a discrete energy spectrum at variance to what happens in an asymptotically
flat Universe. Moreover, the Bohr-Sommerfeld quantization method can be applied 
in order to determine the energy levels and compared with the exact results.
	
		\subsubsection{The $\ell=0$ case}
	The $\ell=0$ case is first considered as some analytical calculations 
can be performed. The two turning points are $r^\star_-=0$ and 
$r^\star_+=R\arctan{\left(\sqrt{\frac{\omega^2R^2-2}{2}}\right)}$ and the 
potential takes the following form $V_0(r^\star)=\frac{\sqrt{2}}{R\cos{\left(\frac{r^\star}{R}\right)}}+\delta(r^\star)$,
where a Dirac distribution was included so as to take into account that the 
negative $r^\star$ region is forbidden. When one of the walls is infinitely 
steep, as is the case here, the Bohr-Sommerfeld quantization rule is slightly 
modified because of the new boundary conditions for the wave function \cite{brack}. 
The quantization rule reads in this case
$
	W_0=\left(2n+\frac{3}{2}\right)\pi.
$
To account for the distribution involved, a more rigorous 
definition of this quantity shall be used~:
\begin{equation}
	W_0=2\lim_{\varepsilon\rightarrow0}\displaystyle\int_{r^\star_-+\varepsilon}^{r^\star_+-\varepsilon}\sqrt{\omega^2-V^2(r^\star)}dr^\star.
\end{equation}

We finally obtain the following value for the classical action 
$W_0=\pi\left(\omega{R}-\sqrt{2}\right)$ which yields to the following spectrum 
for the $\ell=0$ case~:
\begin{equation}
\left(\omega_{n,0}\right)_{n\in{\mathbb{N}}}=\left(\frac{2n+\sqrt{2}+3/2}{R}\right)_{n\in{\mathbb{N}}}.
\end{equation}
The spectrum dependence with respect to the integer $n$ is well 
recovered and, as a consequence, so are the energy gaps. Nevertheless,
the zero point energy slightly differs
from the accurate one~: the semi-classical approach provides 
$\omega_{0,0}=(3/2+\sqrt{2})/R\simeq2.91/R$ instead of $3/R$. This shift 
is due to the specific behaviour of the potential at 
$r^\star=0$ leading to a high departure from 
any quadratic approximation. It 
is important to mention that if an infinite wall was not added, the 
quantization rule would follow Eq. (\ref{bohr}), yielding to an energy 
spectrum
%~:
%\begin{equation}
%		\left(\omega_{n,0}\right)_{n\in{\mathbb{N}}}=\left(\frac{2n+\sqrt{2}+1}{R}\right)_{n\in{\mathbb{N}}}
%\end{equation}
which  mis-estimates the zero point energy by $20 \%
$
because the potential, if viewed as a function of the two variables $(r^\star,\ell)$, is not continuous at 
$(0,0)$.

	\subsubsection{The $\ell\neq0$ case}
For $\ell>0$, the analytical computation of $W_0(\omega)$ is not possible and 
numerical investigations have to be performed. To evaluate the energy levels,
the equation 
\begin{equation}
	W_0(\omega,\ell)-(2n+1)\pi=0
	\label{eq-level}
\end{equation}
is numerically solved. $W_0$ has to be rewritten in a more convenient way 
using the $y=\tan{(r^\star/R)}$ coordinate~:
\begin{equation}
	W_0=2\displaystyle\int_{y_-}^{y^+}\frac{1}{1+y^2}\sqrt{\omega^2R^2-\left(1+y^2\right)\left(\frac{\ell(\ell+1)}{y^2}+2\right)}dy
\end{equation}
where $y_{\pm}$ are the two turning points, determined in appendix \ref{app-b},
\begin{equation}
	y_{\pm}=\frac{1}{2}\sqrt{\omega^2R^2-2-\ell(\ell+1)\pm\sqrt{\left(2+\ell(\ell+1)-\omega^2R^2\right)^2-8\ell(\ell+1)}}.
\end{equation}

\begin{figure} 
	\begin{center}
		\includegraphics[scale=0.7]{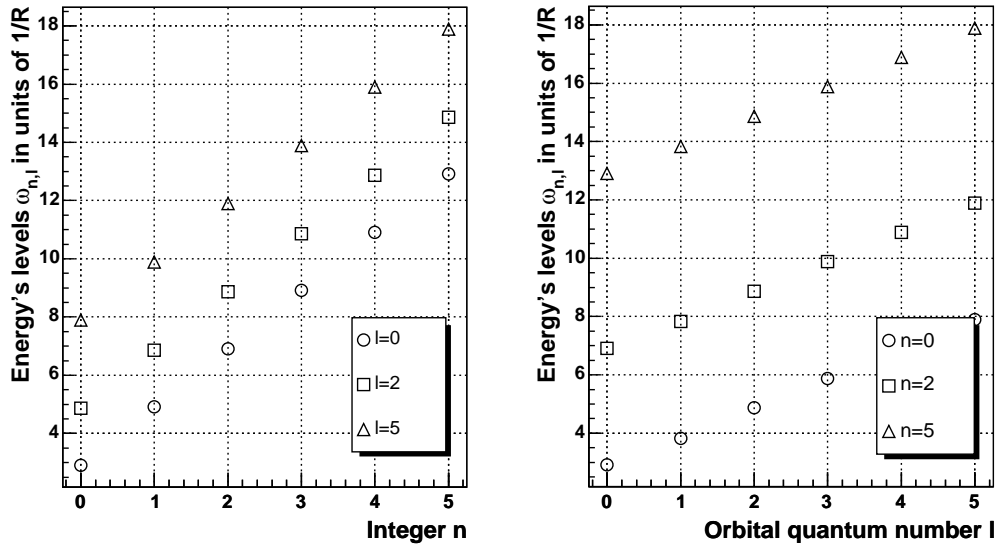}
		\caption{{\it{left panel~:}} Energy levels as a function of the 
		integer $n$. {\it{Right panel~:}} Energy levels as a function of the 
		orbital quantum number $\ell$. All the energies are expressed in units
		 of $R^{-1}$.}
		\label{level-num}
	\end{center}
\end{figure}

\begin{figure} 
	\begin{center}
		\includegraphics[scale=0.7]{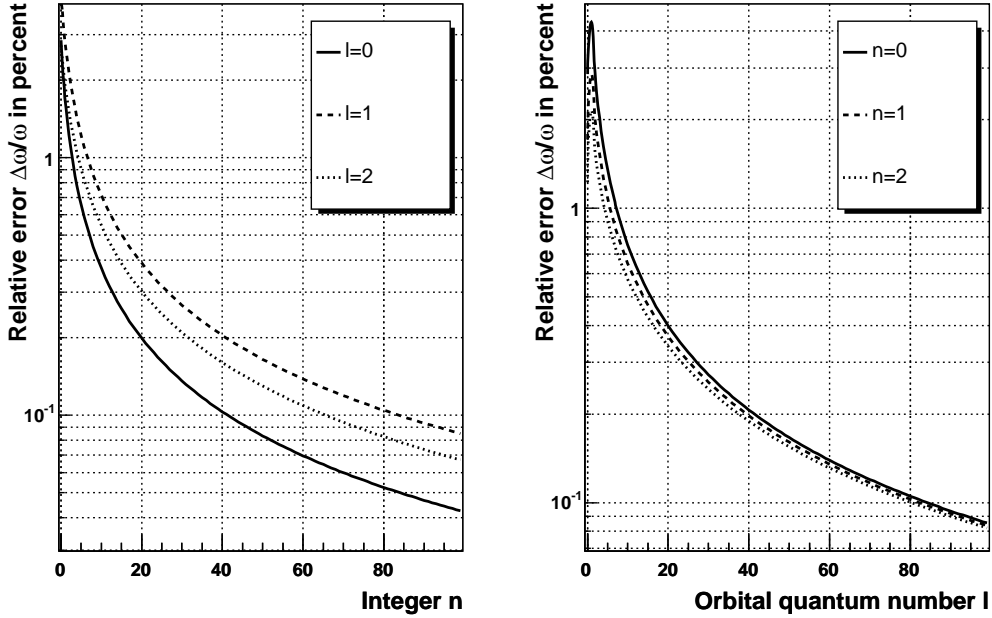}
		\caption{{\it{left panel~:}} Total relative error $\frac{\Delta\omega}{\omega}$ as 
		a function of the integer $n$. {\it{Right panel~:}} Total relative error 
		$\frac{\Delta\omega}{\omega}$ as a function of the orbital quantum number $\ell$.}
		\label{level-error}
	\end{center}
\end{figure}

The numerical results are presented on Fig.~\ref{level-num} and 
Fig.~\ref{level-error} where the energy levels $\omega_{n,\ell}$ and the relative 
errors $\Delta\omega/\omega$ are displayed respectively as a function of the integer 
$n$ and as a function of the orbital quantum number $\ell$. Although slightly underestimated, the energy levels are well 
recovered and the precision of this semi-classical approach increases with $n$ 
and $\ell$~: the total relative error remains below $5\%
$. For a 
vanishing orbital quantum number, the result of the previous analytical 
calculation is recovered, as it should be. When $\ell\neq0$, 
the energy levels as a function of the angular momentum are well described by 
a linear function with a $\ell-$intercept equal to 
$(2n+3)/R$. Recovering the correct zero-point energy by taking the
$\ell\to0$ limit lights up the particular behaviour of the potential for a vanishing angular 
momentum and confirms the prescription used in the previous section.

There are two possible sources of error in those numerical computations. The first one 
is due to numerical approximations and was checked to
remain totally negligible. The second one is related to the Bohr-Sommerfeld 
quantization scheme which gives exact results up to the second order in the 
potential only. This error decreases both with $n$ and with the orbital quantum number 
$\ell$~: for a fixed value of $n$, not only does the energy levels increase with 
$\ell$ but also does the minimum of the potential 
$V_{min}=\left(\sqrt{2}+\sqrt{\ell(\ell+1)}\right)R^{-1}$. This result is expected as 
particles with higher energies are 
closer to ``classicality''.

\section{Greybody factors for Schwarzschild black holes}
\label{grey}

For massless scalars, the flux emitted by an evaporating black hole 
can be written  as \cite{hawking}~:
\begin{equation}
	\frac{d^2N}{d\omega{dt}}=\frac{4\pi\sigma_g(\omega)\omega^2}{e^{\frac{\omega}{T_{BH}}}-1}.
	\label{flux}
\end{equation}
The non trivial part mostly lies in the determination of the greybody factors 
$\sigma_g(\omega)$ (related with the probability for an emitted particle to escape 
form the gravitational potential) whose 
computation involves the resolution of the KG equation, given by Eq. (\ref{kg}),
in the Schwarzschild background. The tunnel probability 
$\left|A_\ell\right|^2$ \cite{page} yields to the greybody factors via the optical theorem 
\cite{sakurai}~:
\begin{equation}
	\sigma_g(\omega)=\displaystyle\sum_\ell\frac{(2\ell+1)\pi}{\omega^2}\left|A_\ell\right|^2.
	\label{optical}
\end{equation}

Unlike the AdS case, the intricate shape of the metric function describing Schwarzschild black 
holes does not allow for an analytical resolution of Eq. (\ref{kg}). As a consequence, all the 
evaluations of the greybody factors, from usual Schwarzschild black holes to $D$-dimensional 
Gauss-Bonnet ones \cite{page,harris,kgb,kgb-gb}, are based on numerical investigations; thought 
some analytical methods can be applied in the high and low energy regimes \cite{kanti}. However, 
using the WKB method, some approximated results can be obtained without big numerical codes. 
The aim of this study is to derive the greybody factors for 
4-dimensional Schwarzschild black holes at the WKB order and to compare them
with the exact results.

	\subsection{WKB wave function and tunnel probability}
\begin{figure} 
	\begin{center}
		\includegraphics[scale=0.6]{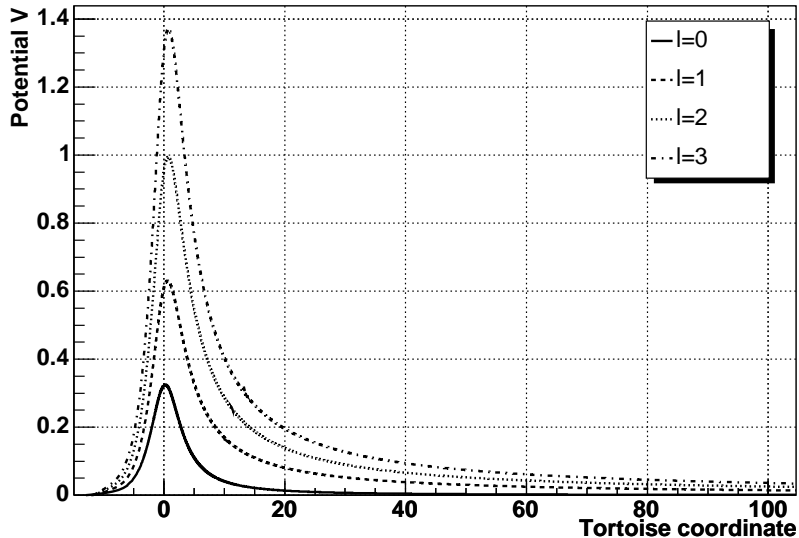}
		\caption{Potential seen by a scalar field propagating in Schwarzschild space-time}
		\label{bh-pot} 
	\end{center}
\end{figure}

For a black hole with a horizon radius $r_H$, we consider the metric~:
\begin{equation}
	ds^2=\left(1-\frac{r_H}{r}\right)dt^2-\frac{dr^2}{1-\frac{r_H}{r}}-r^2d\Omega^2.
\end{equation}
In the tortoise coordinates system $(t,r^\star)$, particles are described by a wave 
function $U$ 
evolving in the potential $V$ (see Fig. \ref{bh-pot})~:
\begin{equation}
	\begin{array}{l}
	
V^2_\ell(r)=\left(1-\frac{r_H}{r}\right)\left(\frac{\ell(\ell+1)}{r^2}+\frac{r_H}{r^3}\right)~;~
		r^\star=r+r_H\ln{\left(r-r_H\right)}.
	\end{array}
	\label{pot-form-bh}
\end{equation}
As it can be seen from Fig. 
\ref{bh-pot}, this potential tends to zero when $r^\star\to\pm\infty$, making the use of the 
optical theorem valid. We define $r^\star_{\mathrm{max}}$ as the tortoise coordinate
corresponding to the maximum value $V_{\mathrm{max}}$.

To determine the wave function, the "no-incoming wave at 
spatial infinity" hypothesis is used as there are no sources of particles far from the black 
hole. For particles with energies always greater than the potential, the wave function 
 reads $U(r^\star)={A}_{out}\exp{\left(i\int^{r^\star}_{r^\star_0}p(x)dx\right)}$ with 
$p=\sqrt{\omega^2-V^2}$. For particles with energies lower than $V_{\mathrm{max}}$, the space is 
divided into three regions $\mathcal{A}$, $\mathcal{B}$ and $\mathcal{C}$, corresponding 
respectively to $]-\infty,r^\star_-]$, $[r^\star_-,r^\star_+]$ and $[r^\star_+,+\infty[$, with 
$(r^\star_-,r^\star_+)$ representing the two turning points. With the above boundary conditions, 
the wave function admits the following asymptotic regime~: 
${U}(r^\star)\simeq{A}_{in}\exp{\left(-i\omega{r}^\star\right)}+{A}_{out}\exp{\left(i\omega{r}^\star\right)}$ when 
$r^\star\rightarrow-\infty$ and ${U}(r^\star)\simeq{B}_{out}\exp{\left(i\omega{r}^\star\right)}$ 
when $r^\star\rightarrow+\infty$. The WKB wave function must therefore be read as~:

\begin{equation}
	{U}(r^\star)=\left\{\begin{array}{ll}
					\frac{1}{\sqrt{p(r^\star)}}\left({A}_{in}e^{-i\int_{r^\star_1}^{r^\star}{p(x)dx}}+{A}_{out}e^{i\int_{r^\star_1}^{r^\star}{p(x)dx}}\right) & ~~\mathrm{in~region}~\mathcal{A} \\
					\frac{A_{eva}}{\sqrt{p'(r^\star)}}e^{-\int_{r^\star_1}^{r^\star}{p'(x)dx}} & ~~\mathrm{in~region}~\mathcal{B} \\
					\frac{B_{out}}{\sqrt{p(r^\star)}}e^{i\int_{r^\star_2}^{r^\star}{p(x)dx}} & ~~\mathrm{in~region}~\mathcal{C}
				  \end{array}\right.
\end{equation}
with $p$ and $p'$ defined as in section \ref{wkb-part}. The above solutions show that particles 
in region $\mathcal{A}$ are described by both incoming waves, {\it{i.e}} particles emitted by the 
black hole, and outgoing ones, {\it{i.e}} particles reflected on the potential barrier, whereas 
they are only described by outgoing waves in region $\mathcal{C}$. In region $\mathcal{B}$, 
particles are only described by evanescent waves as there are no sources of particles at spatial 
infinity. 
Using the procedure described in appendix \ref{app-c} to match the different solutions at the two 
turning points, it is found that $A_{in}=e^{-i\pi/2}A_{out}$ and $B_{out}=e^{-\tau}A_{out}$ with 
$\tau=\int^{r^\star_+}_{r^\star_-}p'(x)dx$. This approach leads to a new understanding of
the evaporation process.
			
Unlike the AdS case, it is 
not possible to inverse the relation $r^\star=f(r)$ and an explicit expression 
of the potential as a function of $r^\star$ cannot be found. However, the determination 
of the two 
turning points $r^\star_\pm$ as well as $r^\star_{\mathrm{max}}$ can be 
performed using $r$ as a variable instead of $r^\star$. The tunnel probability 
is given by the ratio of the outgoing flux 
${\mathcal{F}}_U=i(U\partial_{r^\star}{U}^\dag-U^\dag\partial_{r^\star}{U})$ at 
spatial infinity to the one at the event horizon of the black hole. Once 
reexpressed using the usual radial coordinate, the tunnel probability 
$\left|\tilde{A}_\ell\right|^2$ is given by~:
\begin{equation}
	\left|\tilde{A}_\ell\right|^2=\left\{\begin{array}{ll}
			1 & ~\mathrm{if}~~\omega>V_{\mathrm{max}} \\
			e^{-2\int_{r_-}^{r_+}{\frac{\sqrt{V^2(r)-\omega^2}}{h(r)}dr}} & ~\mathrm{if}~~\omega<V_{\mathrm{max}}
		      \end{array}\right.
\end{equation}
where $r_{\pm}$ are the two turning points and $V_{\mathrm{max}}$ is obtained 
by computing the 
value of the potential at $r_{\mathrm{max}}$ so that
\begin{equation}
	r_{\mathrm{max}}=\left\{\begin{array}{ll}
			\frac{4}{3}r_H & ~\mathrm{for}~~\ell=0 \\
			 & \\
			\frac{r_H}{4\ell(\ell+1)}\left(\begin{array}{l}
				3\ell(\ell+1)-3 \\
				+\sqrt{9+9\ell^2(\ell+1)^2+14\ell(\ell+1)}
			\end{array}\right) & ~\mathrm{for}~~\ell\neq0.
		    \end{array}\right.
\end{equation}

	Although the maximum value of the potential is analytically determined, 
the two turning points as well as the integral involved in the calculation of 
the tunnel probability have to be numerically computed. The above tunnel probability  has
to be linked with the transmission probability 
$\left|{A}_\ell\right|^2$, which is obtained 
using the flux for the $R$ function 
$\mathcal{F}_R=i(hr^2)({r^{-2}R}\partial_rr^2R^\dag-{r^{-2}R^\dag}\partial_rr^2R)$. 
As $R(r)=U(r)/r$, the usual radial wave function $R$ is composed of incoming 
and outgoing spherical waves with amplitudes equal to the ones of the $U$ 
function and the transmission probability $\left|{A}_\ell\right|^2$ is given by 
$\left|B_{out}/A_{out}\right|^2=\left|\tilde{A}_\ell\right|^2$. The 
WKB tunnel probability $\left|\tilde{A}_\ell\right|^2$ can therefore be directly used
in Eq. (\ref{optical}) to compute the value of the greybody factor.

	\subsection{Results for the cross-section and radiation spectra}
\begin{figure} 
	\begin{center}
		\includegraphics[scale=0.7]{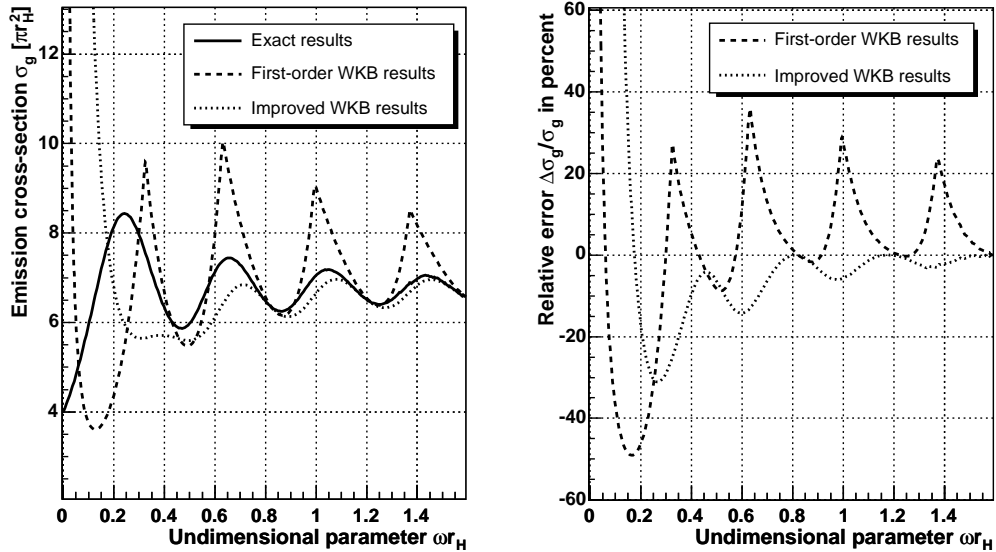}
		\caption{{\it{Left panel~:}} Emission cross-section for a $4-$dimensional 
		Schwarzschild black hole. The solid line corresponds to exact numerical results, 
		the dashed line to the first-order WKB approximation results and the dotted line 
		to the improved WKB results. {\it{Right panel~:}} Relative error $\Delta\sigma_g/\sigma_g$ of the WKB cross-section estimation. The dashed line stands for the first-order WKB error and the dotted one to the improved WKB error.}
		\label{grey-fig} 
	\end{center}
\end{figure}

Figure.~\ref{grey-fig} shows, on the left panel, the WKB emission/absorption 
cross-section alongside the exact one \cite{page}, numerically obtained 
\cite{kgb}, and, on the right panel, the relative error. 
We expect the WKB method to break down twice~: in the
low energy region because
this is where the quantum nature of particles dominates and near
the resonances, {\it{i.e}} $\omega=V_{\mathrm{max}}$, because the two turning 
points are extremely close one to the other. This behaviour can be
seen by looking at the shape of the emission/absorption cross-section. The divergence 
is due to the $\ell=0$ partial wave whose transition probability tends to 
$e^{-4\sqrt{r_H}}$ when $\omega$ tends to zero. On the other hand, the cross-section 
at resonances, occurring when the energy of the field is equal to $V_\ell$, 
is overestimated in the WKB approach (although the spectrum 
is quite well recovered).
The mis-estimation at resonances is not dramatic and leads to regular peaks in 
the relative error whose amplitude does not exceed $35\% 
$. Furthermore, it has only small consequences on the radiation spectrum. 
However, without the help of an exact calculation, the WKB 
approximation seems --as expected-- hazardous to provide reliable results in the low-energy region.
It is both due to the unavoidable behaviour of the turning points and to the huge steps for the
tortoise coordinate numerically occuring close to the event horizon.

In spite of this drawback, the WKB techniques provides good results in the 
intermediate and high energy regimes. First, the high energy optical limit is 
well recovered, as it can be seen on Fig. \ref{grey-fig}. Moreover, it was 
demonstrated in Ref. \cite{cornell} that if the transmission coefficient 
$\left|{\mathcal{A}}_\ell\right|^2$ tends to unity when $\omega$ tends to 
infinity for all the partial waves, then the emission/absorption cross-section
should tend to the optical limit when $\omega\to+\infty$. Our WKB transmission 
coefficient is in agreement 
with the above conditions and the cross-section has to tend to 
$\sigma^{({\mathrm{opt}})}_g=27\pi{r}^2_H/4$. In the intermediate 
energy regime, the relative error admits an average value of $10\%
$ and decreases for higher values of the energy.

Those results could be improved using higher order semi-classical
 expansions. A third-order WKB tunnel probability has been derived by Iyer \& Will and extended 
 to the sixth order by Konoplya (see  Ref.~\cite{qnm}) to derive the quasi-normal frequencies of
  black holes. It is optimised for resonant scattering near the 
  top of the potential barrier and allows for a non-vanishing reflexion coefficient for 
  particles with an energy greater than the maximum of the potential. This should allow to 
  avoid the spiky structure at resonances. The
  transmission coefficient, successfully used to derive the greybody factors for gravitons emitted 
  by $D-$dimensional Schwarzschild black holes \cite{cornell}, takes the following form
\begin{equation}
	\left|\mathcal{A}_\ell\right|^2=\frac{1}{1+e^{2i\pi\left(\nu+1/2\right)}}~,~\nu+1/2=i{p}^2(r_{\mathrm{max}})\left(2\frac{d^2{p}^2}{d{r^\star}^2}(r_{\mathrm{max}})\right)^{-1/2}.
	\label{trans-seuil}
\end{equation}

The greybody factors obtained with the improved WKB tunnel probability are displayed on 
Fig.~\ref{grey-fig} where it can be seen that the breakdown of the approximation at the top 
of the potential barrier is removed. 
 However, the infrared divergence cannot be avoided with this new technique, as is the case for 
 gravitons \cite{cornell}.

\begin{figure} 
	\begin{center}
		\includegraphics[scale=0.6]{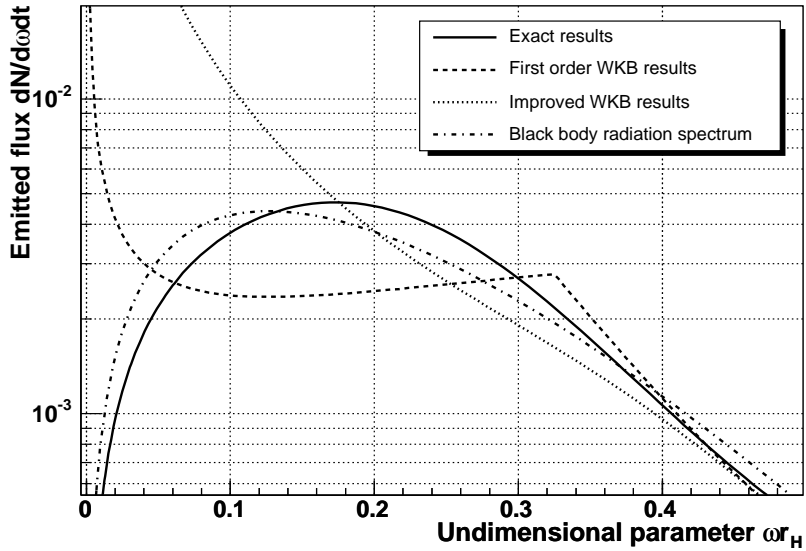}
		\caption{Scalar radiation spectrum from a $4-$dimensional Schwarzschild black hole.
		 The solid line corresponds to exact numerical results, the dashed one to the 
		 first-order WKB approximation results, the dotted line to the improved WKB 
		 results and the dashed-dotted one to a black body radiation spectrum with the 
		 high energy limit of the cross-section.}
		\label{flux-fig} 
	\end{center}
\end{figure}
	
The emission/absorption cross-section does not 
correspond to any physical quantity. The meaningful flux,
or radiation spectrum, is displayed on Fig.~\ref{flux-fig}. 
As it can clearly be seen, both WKB approximations allow for a good 
estimation of the high energy tail of the distribution whereas the low 
energy one is mis-estimated due to the infrared divergence. 
It should be pointed out that the WKB technique for 
the estimation of greybody factors is more of theoretical interest~: it allows a 
first and fast estimation of the 
emission/absorption cross-section in both the intermediate and high energy 
regime (which is highly relevant to understand the consequences of the possibly intricate metric),
 but does not provide any real improvement in the determination of the 
"experimental"  spectrum.

\section{Scalar fields in SAdS space-times}
\label{sads}

	\subsection{Shape of the potential}
	Fields in SAdS space-times have been intensively studied for quasi-normal modes 
estimation (see Ref. \cite{qnm,qnm-ads} and references therein).
Nevertheless, the general behaviour of scalar fields in those space-times has not been 
investigated and this study 
is carried out at the semi-classical order hereafter. SAdS space-times are described by a 
Schwarzschild-like metric with 
\begin{equation}
	h(r)=1-\frac{2M}{r}+\frac{r^2}{R^2}.
\end{equation}
The black hole horizon radius corresponds to the positive real root of $h(r)$~: 
$2M=r_H(1+r^2_H/R^2)$. The potential seen by a scalar field and the tortoise coordinate are given by
\begin{equation}
	\begin{array}{l}
	V^2_\ell(r)=\left(1-\frac{2M}{r}+\frac{r^2}{R^2}\right)\left(\frac{\ell\left(\ell+1\right)}{r^2}+\frac{2M}{r^3}+\frac{2}{R^2}\right)~; \\
		r^\star=R^2\left[\frac{r_H}{3r^2_H+R^2}\ln{\left(\frac{r-r_H}{\sqrt{r^2+rr_H+r^2_H+R^2}}\right)}\right. \\
	~~~~~~~~~~~\left.+\frac{3r^2_H+2R^2}{\left(3r^2_H+R^2\right)\sqrt{3r^2_H+4R^2}}\arctan{\left(\frac{2r+r_H}{\sqrt{3r^2_H+4R^2}}\right)}\right].
	\end{array}
\end{equation}

\begin{figure} 
	\begin{center}
		\includegraphics[scale=0.6]{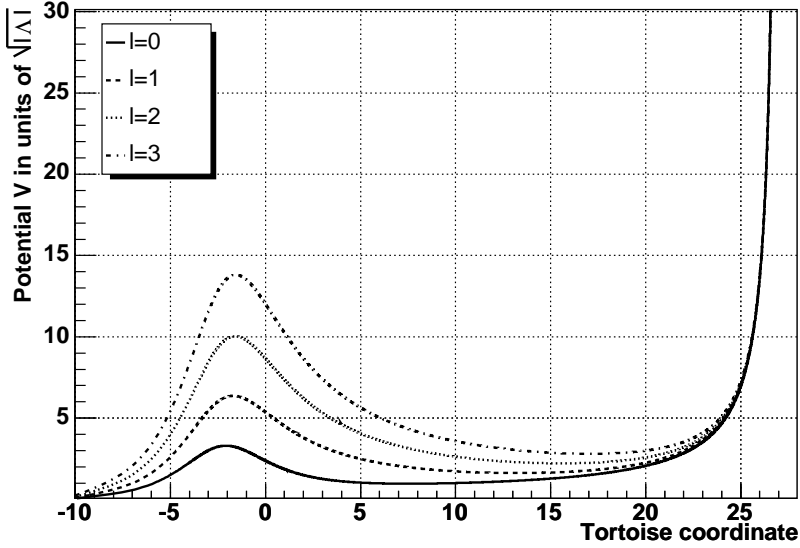}
		\caption{The potential in units of $\sqrt{\left|\Lambda\right|}$ seen by a 
		massless scalar field in SAdS space-time as a function of the tortoise coordinate 
		for $r_H=0.1$ and for $\eta\ll1$.}
		\label{r-star-pot} 
	\end{center}
\end{figure}

The shape of the potential strongly depends upon  $\eta=r_H/R$. If $\eta$ 
is much smaller than one, the black hole horizon radius is 
smaller than the characteristic AdS curvature radius, while if $\eta$ is much greater than one, 
the horizon radius becomes greater than the curvature radius. The potential is depicted on 
Fig.~\ref{r-star-pot} for $\eta\ll1$ and is a mixing of 
the potentials in Schwarzschild space-time and in AdS space-time~: when 
$r^\star\to-\infty$, {\it{i.e.}} $r\to{r}_H$, the potential falls down to zero because of the 
event horizon whereas it diverges to $+\infty$ when $r^\star\to{r}^\star_{\mathrm{SAdS}}$, 
{\it{i.e.}} $r\to+\infty$, where 
$r^\star_{\mathrm{SAdS}}=\frac{R^2(3r^2_H+2R^2)}{(3r^2_H+R^2)\sqrt{3r^2_H+4R^2}}\frac{\pi}{2}$. 
It also exhibits a finite barrier -because of the presence of the black hole-
for any value of the orbital quantum number $\ell$, which leads to the 
appearance of a local maximum $V^{(\ell)}_{\mathrm{max}}$ and minimum 
$V^{(\ell)}_{\mathrm{min}}$, respectively located at $r^\star_{\mathrm{max}}$ and 
$r^\star_{\mathrm{min}}$. 

When $\eta\gg1$, the finite potential barrier is present only at high 
values of the angular momentum~: as examples, the barrier appears at $\ell\simeq10$ for $\eta=1$ 
and at $\ell\simeq10^6$ for $\eta=2$. In the following, we will focus on the high hierarchy 
case ($\eta\ll1$) since it corresponds to the most physically interesting one.

	\subsection{Stationary states solutions}

The solution being developped in stationary states 
$\psi(r^\star,t)=\psi_r(r^\star)e^{-i\omega{t}}$, the equation of motion is given by 
Eq. (\ref{schro}) where $\omega$ is a real and positive parameter. The WKB solutions depends on the 
energy\footnote{Calling {\it{energy}} the frequency parameter $\omega$ was meaningful in
the previously studied space-times as it was corresponding to the energy of particles for 
observers located at $r=0$ in AdS Universe or at spatial infinity in Schwarzschild ones. 
Although such {\it{flat}} observers do not exist in SAdS background, we keep calling 
{\it{energy}} the frequency parameter to avoid using too many variables.} $\omega$ of the field 
and the problem has to be divided into three energy regimes, defined by the number of classical 
turning points~: i) $\omega<V^{(\ell)}_{\mathrm{min}}$ (Regime I), one turning point $r_-$
located before $r^{(\ell)}_{\mathrm{max}}$ ; ii)
$V^{(\ell)}_{\mathrm{min}}<\omega<V^{(\ell)}_{\mathrm{max}}$ (Regime II), three turning points, 
$r_\pm$ corresponding to the potential barrier induced by the black hole and $r_0$ corresponding 
to the infinite barrier at spatial infinity ; iii) $V^{(\ell)}_{\mathrm{max}}<\omega$ (Regime III),
one turning point $r_0$ which also corresponds to the infinite potential barrier. The boundary 
conditions, although depending on the considered quantity (quasi-normal frequencies, 
greybody factors, {\it etc}.), should read as follows in the most general case~:

\begin{equation}
	\psi_r(r^\star)\to\left\{\begin{array}{ll}
			\frac{A_{in}}{\sqrt{\omega}}e^{-i\omega{r}^\star}+\frac{A_{out}}{\sqrt{\omega}}e^{i\omega{r}^\star} & {\mathrm{when}}~r^\star\to-\infty \\
			0 & {\mathrm{when}}~r^\star\to{r}^\star_{\mathrm{SAdS}}.
		\end{array}\right.
\end{equation}

Considering first the {\emph{one-turning-point}} case, the WKB wave functions has to be 
considered in two regions~: i) $r<r_{-/0}$ (Region $\mathcal{A}$) where the solutions are 
progressive waves and ii) $r>r_{-/0}$ (Region $\mathcal{B}$) where the solutions are evanescent 
and exponentially divergent waves. With proper boudary conditions,
the WKB wave functions are obtained by the matching procedure at $r<r_{-/0}$ described in 
appendix \ref{app-c}~:
\begin{equation}
	\psi_r(r^\star)=\left\{\begin{array}{ll}
			\frac{A_{out}}{\sqrt{p(r^\star}}\left(e^{i\int^{r^\star}{p}(x)dx}-ie^{-i\int^{r^\star}{p}(x)dx}\right) & {\mathrm{in~Region}}~{\mathcal{A}} \\
		\frac{A_{out}}{\sqrt{p'(r^\star}}e^{-i\frac{\pi}{4}-\int^{r^\star}_{r^\star_{-/0}}{p'}(x)dx} &
{\mathrm{in~Region}}~{\mathcal{B}}.
		\end{array}\right.
\end{equation}

For the {\emph{three-turning-points}} case, the WKB solutions are considered in four regions~: i) 
$r<r_{-}$ (Region $\mathcal{A}$) where the solutions are progressive waves ; ii) $r_-<r<r_+$ 
(Region $\mathcal{B}$) where the solutions are evanescent and exponentially divergent waves ; iii)
$r_+<r<r_0$ (Region $\mathcal{C}$) where the solutions are also progressive waves and iiii) 
$r>r_0$ (Region $\mathcal{D}$) where the solutions are only evanescent waves, taking into account
the boundary conditions at $r^\star_{\mathrm{SAdS}}$. Applying the matching procedure, the WKB 
solutions are found to be~:
\begin{equation}
	\psi_r(r^\star)=\left\{\begin{array}{ll}
			\frac{A_{out}}{\sqrt{p(r^\star}}\left(e^{i\int^{r^\star}{p}(x)dx}-ie^{-i\int^{r^\star}{p}(x)dx}\right) & {\mathrm{in~Region}}~{\mathcal{A}} \\
			\frac{A_{out}}{\sqrt{p'(r^\star}}e^{-i\frac{\pi}{4}-\int^{r^\star}_{r^\star_-}{p'}(x)dx} & {\mathrm{in~Region}}~{\mathcal{B}} \\
			\frac{A_{out}e^{-\tau}}{2\sqrt{p(r^\star}}\left(e^{i\int^{r^\star}{p}(x)dx}-ie^{-i\int^{r^\star}{p}(x)dx}\right) & {\mathrm{in~Region}}~{\mathcal{C}} \\
			\frac{A_{out}e^{-\tau}}{2\sqrt{p'(r^\star}}e^{-i\frac{\pi}{4}-\int^{r^\star}_{r^\star_0}{p'}(x)dx} & {\mathrm{in~Region}}~{\mathcal{D}}.
		\end{array}\right.
\end{equation}
where $\tau=\int^{r_+}_{r_-}p'(x)dx$. 
This strange behaviour in Region  $\mathcal{B}$ arises because we have assumed that the solutions 
can be 
decomposed into stationary state and because of the semi-classical anstaz. When considering the 
shape of the potential, it is clear that the states in Region $[r^{(\ell)}_{\mathrm{max}},+\infty]$
cannot be stationary states but quasi-stationary ones, as tunneling transition from Region 
$\mathcal{C}$ to Region $\mathcal{A}$ is possible but not taken into account in a first order 
semi-classical expansion.

The radial density probability for scalar particles can be inferred from the above stationary 
state solutions and takes the following form~:
\begin{equation}
	\rho(r^\star)=i\left[\psi_r(r^\star)e^{-i\omega{t}}\partial_t\left(\psi^\dag_r(r^\star)e^{i\omega{t}}\right)-\psi^\dag_r(r^\star)e^{i\omega{t}}\partial_t\left(\psi_r(r^\star)e^{-i\omega{t}}\right)\right]=2\omega\psi^\dag_r(r^\star)\psi_r(r^\star).
\end{equation}
In the three-turning-points case, the density probability therefore reads as
\begin{equation}
	\rho(r^\star)=2\omega\left|A_{out}\right|^2\left\{\begin{array}{ll}
				\frac{2}{p(r^\star)}\left[1+\cos{\left(2\int^{r^\star}{p}(x)dx+\frac{\pi}{2}\right)}\right] & {\mathrm{in~Region}}~{\mathcal{A}} \\
				\frac{1}{p'(r^\star)}\exp{\left(-2\int^{r^\star}_{r^\star_-}{p'}(x)dx\right)} & {\mathrm{in~Region}}~{\mathcal{B}} \\
				\frac{e^{-2\tau}}{2p(r^\star)}\left[1+\cos{\left(2\int^{r^\star}_{r^\star_+}{p}(x)dx+\frac{\pi}{2}\right)}\right] & {\mathrm{in~Region}}~{\mathcal{C}} \\
				\frac{e^{-2\tau}}{4p'(r^\star)}\exp{\left(-2\int^{r^\star}_{r^\star_0}{p'}(x)dx\right)} & {\mathrm{in~Region}}~{\mathcal{D}}.
		\end{array}\right.
\end{equation}
\begin{figure}
	\begin{center}
		\includegraphics[scale=0.5]{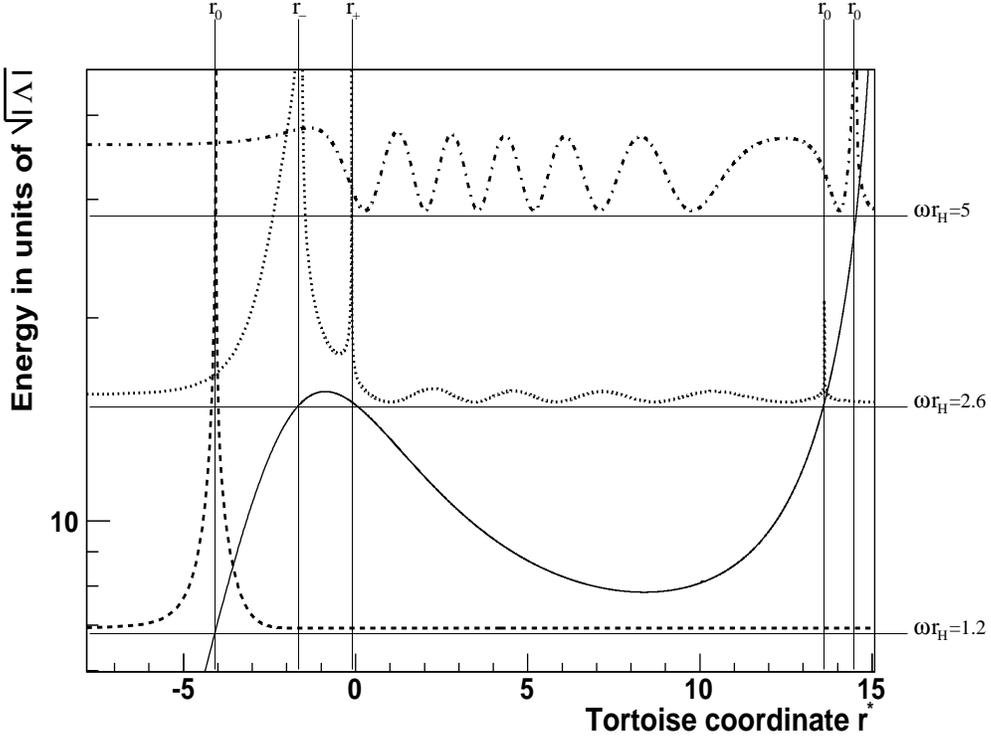}
	\end{center}
	\caption{Probability densities as functions of 
	$r^\star$ alongside the $\ell=6$ scalar potential (solid line) for 
	$\eta=10$~: the dashed line stands for 
	$\omega<V^{(\ell)}_{\mathrm{min}}$, the dotted line for 
	$V^{(\ell)}_{\mathrm{min}}<\omega<V^{(\ell)}_{\mathrm{max}}$  and 
	the dotted-dashed line for  $\omega>V^{(\ell)}_{\mathrm{max}}$. 
	Each probability is displayed at its energy $\omega_i$, the 
	horizontal line at $\omega=\omega_i$ corresponding to $\rho=0$. 
	The turning-points are 
	shown and the energy can be inferred either 
	in units of $\sqrt{\left|\Lambda\right|}$ on the left $y-$axis or in units 
	of $r^{-1}_H$ on the right.}
	\label{dens-prob}
\end{figure}
This probability is displayed on Fig.~\ref{dens-prob} for three different cases.
As expected, at the turning points, 
the probability density diverges because of the $p^{-1}(r^\star)$ factor, which is a direct 
consequence of the breakdown of the WKB approximation. Since the potential rapidly falls down 
to zero, the radial probability 
density rapidly tends to the coherent superposition of ingoing and 
outgoing free particles, {\it{i.e.}} plane waves, with the same amplitude
$
	\lim_{r^\star\to-\infty}{\rho(r^\star)}=\frac{1}{2}\left[1+\cos{\left(2\omega{r}^\star+\frac{\pi}{2}+\varphi(r^\star_{\mathrm{ini}})\right)}\right],
$
where $\varphi$ is an additional phase depending on the initial position chosen for integration. 
The minimum and maximum energy cases can be divided into two parts~: before the turning point, 
the probability density is described by interferences between ingoing and outgoing modes, whereas 
after the turning point the probability density is a decreasing exponential which falls down to 
zero at $r^\star=r^\star_{\mathrm{SAdS}}$. For the intermediate energy case, the space is 
separated into four regions~: for $r<r_-$ and $r_+<r<r_0$, the probability density takes the form 
of interferences between ingoing and outgoing modes whereas it takes the form of a decreasing 
exponential in the $r_-<r<r_+$ and $r>r_0$ regions and falls down to zero at 
$r^\star=r^\star_{\mathrm{SAdS}}$.

Some conclusions about the black hole evaporation process with a negative cosmological
constant can be inferred in this framework. As in SdS space-time, the evaluation of the 
transmission coefficient critically depends upon the position of the observer. If the observer 
is located at spatial infinity, the transmission coefficient falls down to zero because of the 
total reflexion at $r_0$. 
However, this can be avoided by considering an observer near the local minimum, 
which is always possible in the case of a high hierarchy. Particles with an energy lying 
in Regime I are then totally reflected at $r_-$ and the observer should not see any particle with an 
energy lower than $V^{(0)}_{\mathrm{min}}$ coming from the black hole. In addition, for quanta 
with $\omega>V^{(0)}_{\mathrm{min}}$, only modes with angular momentum lower than $m$, so that 
$V^{(m-1)}_{\mathrm{min}}<\omega<V^{(m)}_{\mathrm{min}}$, can reach the observer. 
For $\omega<V^{(\ell)}_{\mathrm{max}}$, the transmission 
coefficient is given by $e^{-2\tau}$ while it becomes equal to one for particles with 
$\omega>V^{(\ell)}_{\mathrm{max}}$. The qualitative features of the evaporation process of SAdS 
black holes can therefore be captured. It should, of course, be pointed out that a realistic study 
of this phenomenon has to be carried out dynamically, involving a full resolution of the KG equation 
while keeping the time dependence.

	\subsection{Energy levels and bandwidths}

Because of the shape of the SAdS 
potential, the density of states for scalar fields is smooth and
continuous with some resonances at specific energies, lying in the 
$[V_{\mathrm{min}},V_{\mathrm{max}}]$ range, with bandwidths given by the tunnel probabilities. 
This should lead to stationary 
states when the potential barrier becomes either infinitely large or infinitely high. 
The aforementioned conditions can be obtained for high angular momentum quantum numbers as well
as for a vanishing $\eta$ ratio. 
Working at the semi-classical order, the energy levels are computed by 
solving numerically the following algebraic equation~:
\begin{equation}
	2\displaystyle\int^{r_0}_{r_+}\frac{\sqrt{\omega^2_{n,\ell}-V^2_\ell(r)}}{h(r)}dr-\left(2n+1\right)\pi=0,
	\label{equ-lev-sads}
\end{equation}
where $\omega^2_{n,\ell}$ stands for the unknown quantity. Unlike the AdS case, the usual radial 
coordinate $r$ is prefered over the tortoise one. The numerical investigations show that, 
at the semi-classical order, the product 
$\omega_{n,\ell}\times{R}$ only depends on the ratio $\eta$, $(n,\ell)$ being fixed. 
Consequently, the energy levels can be written in a very convenient way~:
$
	\omega_{n,\ell}\times{R}\approx f\left(n,\ell\right)\times{g}(\eta),
$
where $f$ and $g$ are two functions which respectively depend on $\left(n,\ell\right)$ and $\eta$.
The bandwidth $\Gamma$ of the resonances is 
given by the product of the tunnel probability $\left|A_\ell\right|^2$ with the number of 
collisions at the turning point $r_+$~:
$
	\Gamma_{n,\ell}=\omega_{n,\ell}\times\left|A_\ell\right|^2,
$
with $\left|A_\ell\right|^2$ given by
\begin{equation}
	\left|A_\ell\right|^2=\left\{\begin{array}{l}
			\exp{\left(-2\displaystyle\int^{r_+}_{r_-}\frac{\sqrt{V^2_{\ell}-\omega^2_{n,\ell}}}{h(r)}dr\right)} \\
		\left(1+\exp{\left(-2\pi{p}^2(r_{\mathrm{max}})\left(2\frac{d^2{p}^2}{d{r^\star}^2}(r_{\mathrm{max}})\right)^{-1/2}\right)}\right)^{-1}.
		\end{array}\right.
\end{equation}
In view of the
results obtained for the greybody factors, the second WKB tunnel probability should be preferred 
as it provides more accurate results. The numerical investigations lead to assume that this 
bandwidth can also be separated 
into a $(n,\ell)$ dependence and a $\eta$ dependence, as in the case of the resonant energy 
position.

\begin{table}
	\begin{center}
		\begin{tabular}{c||c|c|c|c}
			 & $n=0$ & $n=1$ & $n=2$ & $n=3$ \\ \hline\hline
			$\ell=0$ & $2.49-i0.24$ & $4.51-i0.44$ & $6.52-i0.63$ & $8.53-i0.83$ \\ \hline
			$\ell=1$ & $3.82-i0.02$ & $5.82-i0.04$ & $7.82-i0.05$ & $9.82-i0.07$ \\ \hline
			$\ell=2$ & $4.86-i1.6~10^{-3}$ & $6.86-i2.3~10^{-3}$ & $8.86-i3.0~10^{-3}$ & $10.85-i3.7~10^{-3}$ \\ \hline
			$\ell=3$ & $5.88-i9.0~10^{-5}$ & $7.88-i1.2~10^{-4}$ & $9.88-i1.5~10^{-4}$ & $11.87-i1.8~10^{-4}$ \\ \hline
			$\ell=4$ & $6.88-i4.6~10^{-6}$ & $8.88-i6.0~10^{-6}$ & $10.88-i7.3~10^{-6}$ & $12.88-i8.7~10^{-6}$ \\ \hline
			$\ell=5$ & $7.89-i2.3~10^{-7}$ & $9.89-i2.9~10^{-7}$ & $11.88-i3.5~10^{-7}$ & $13.88-i4.1~10^{-7}$ 
		\end{tabular}
		\caption{Complex energy levels $\left(\nu_{n,\ell}\times{R}\right)$ for
		$\eta=5\times10^{-4}$ as functions of the angular momentum quantum number $\ell$
		and the integer $n$ of the Bohr-Sommerfeld quantization rule.}
		\label{sads-lev-4}
	\end{center}
\end{table}

\begin{table}
	\begin{center}
		\begin{tabular}{c||c|c|c|c}
			 & $n=0$ & $n=1$ & $n=2$ & $n=3$ \\ \hline\hline
			$\ell=0$ & $2.51-i0.24$ & $4.54-i0.44$ & $6.55-i0.64$ & $8.56-i0.83$ \\ \hline
			$\ell=1$ & $3.83-i0.027$ & $5.82-i0.042$ & $7.82-i0.057$ & $9.81-i0.071$ \\ \hline
			$\ell=2$ & $4.86-i1.6~10^{-3}$ & $6.86-i2.3~10^{-3}$ & $8.86-i3.0~10^{-3}$ & $10.85-i3.7~10^{-3}$ \\ \hline
			$\ell=3$ & $5.88-i9.0~10^{-5}$ & $7.87-i1.2~10^{-4}$ & $9.87-i1.5~10^{-4}$ & $11.87-i1.8~10^{-4}$ \\ \hline
			$\ell=4$ & $6.88-i4.6~10^{-6}$ & $8.88-i6.0~10^{-6}$ & $10.88-i7.4~10^{-6}$ & $12.88-i8.7~10^{-6}$ \\ \hline
			$\ell=5$ & $7.89-i2.3~10^{-7}$ & $9.89-i2.9~10^{-7}$ & $11.89-i3.5~10^{-7}$ & $13.88-i4.1~10^{-7}$ 
		\end{tabular}
		\caption{Complex energy levels $\left(\nu_{n,\ell}\times{R}\right)$ for
		$\eta=10^{-3}$ as functions of the angular momentum quantum number $\ell$ and the
		integer $n$ of the Bohr-Sommerfeld quantization rule.}
		\label{sads-lev-3}
	\end{center}
\end{table}

\begin{table}
	\begin{center}
		\begin{tabular}{c||c|c|c|c}
			 & $n=0$ & $n=1$ & $n=2$ & $n=3$ \\ \hline\hline
			$\ell=0$ & $2.61-i0.26$ & $4.64-i0.47$ & $6.63-i0.70$ & $8.62-i0.97$  \\ \hline
			$\ell=1$ & $3.82-i0.028$ & $5.79-i0.044$ & $7.77-i0.060$ & $9.74-i0.079$  \\ \hline
			$\ell=2$ & $4.85-i1.7~10^{-3}$ & $6.83-i2.4~10^{-3}$ & $8.81-i3.2~10^{-3}$ & $10.78-i4.1~10^{-3}$  \\ \hline
			$\ell=3$ & $5.87-i9.2~10^{-5}$ & $7.85-i1.2~10^{-4}$ & $9.83-i1.6~10^{-4}$ & $11.80-i2.0~10^{-4}$  \\ \hline
			$\ell=4$ & $6.88-i4.8~10^{-6}$ & $8.86-i6.2~10^{-6}$ & $10.84-i7.7~10^{-6}$ & $12.82-i9.3~10^{-6}$  \\ \hline
			$\ell=5$ & $7.88-i2.4~10^{-7}$ & $9.87-i3.0~10^{-7}$ & $11.85-i3.7~10^{-7}$ & $13.83-i4.4~10^{-7}$ 
		\end{tabular}
		\caption{Complex energy levels $\left(\nu_{n,\ell}\times{R}\right)$ for $\eta=10^{-2}$ as functions of the angular momentum quantum number $\ell$ and the integer $n$
		of the Bohr-Sommerfeld quantization rule.}
		\label{sads-lev-2}
	\end{center}
\end{table}

\begin{table}
	\begin{center}
		\begin{tabular}{c||c|c|c|c}
			 & $n=0$ & $n=1$ & $n=2$ & $n=3$ \\ \hline\hline
			$\ell=0$ & $2.65-i0.75$ & / & / & / \\ \hline
			$\ell=1$ & $3.6-i0.12$ & $5.4-i0.95$ & / & / \\ \hline
			$\ell=2$ & $4.74-i8.6~10^{-3}$ & $6.52-i0.055$ & $8.22-i0.459$ & $9.81-i3.42$ \\ \hline
			$\ell=3$ & $5.77-i5.5~10^{-4}$ & $7.58-i2.8~10^{-3}$ & $9.34-i0.019$ & $11.03-i0.15$ \\ \hline
			$\ell=4$ & $6.79-i3.4~10^{-5}$ & $8.62-i1.5~10^{-4}$ & $10.41-i8.2~10^{-4}$ & $12.15-i5.5~10^{-3}$ \\ \hline
			$\ell=5$ & $7.80-i2.0~10^{-6}$ & $9.65-i8.3~10^{-6}$ & $11.46-i3.9~10^{-5}$ & $13.23-i2.2~10^{-4}$ 
		\end{tabular}
		\caption{Complex energy levels $\left(\nu_{n,\ell}\times{R}\right)$ for
		$\eta=10^{-1}$ as functions of the angular momentum quantum number $\ell$ and the
		integer $n$ of the Bohr-Sommerfeld quantization rule.}
		\label{sads-lev-1}
	\end{center}
\end{table}

The semi-classical characteristics of resonances numerically obtained are presented in 
Tables~\ref{sads-lev-4} to \ref{sads-lev-1} for four different values of the ratio 
$\eta=\{0.0005,0.001,0.01,0.1\}$ corresponding to cases where the finite potential barrier 
appears at any multipolar order. The tables give the complex energy 
$\nu_{n,\ell}=\omega_{n,\ell}-i\Gamma_{n,\ell}$ in units of  $R^{-1}$. As the KG equation is 
solved with the ansatz $\Psi=e^{-i\nu{t}}Y^m_\ell(\Omega)R(r)$, the imaginary part of the 
complex energy has to be negative. The positions of resonances 
tend to the pure AdS energy spectrum when $\eta$ tends to zero for all angular momenta except 
when $\ell=0$. This is not surprising since, for 
SAdS space-times, an infinite steepness of the wall never arises, even for arbitrarily small
black 
holes~: the transition from AdS to SAdS space-times is singular. 

The bandwidth of the resonances behaves as expected~: on the one hand, either a decrease of $\eta$ or 
an increase of $\ell$ leads to a decrease of the bandwidth since the well-potential is deeper and,
on the other hand, the bandwidth becomes larger for higher values of  $n$ because of the
increasing tunnel probability. Finally, when $\eta\rightarrow 0$, the quasi-stationary states tend to the stationary ones of 
an AdS Universe 
except for a vanishing angular momentum quantum number. The specific behaviour of the
monopole partial waves underlines the singular behaviour of the SAdS scalar potential. 
However, it
cannot be excluded that this arises from the use of semi-classical methods and is only 
an artifact of this approximated scheme.

\section{Conclusions and perspectives}

It was shown in this article that semi-classical methods for the resolution of Schr\"odinger-like
equations can be successfully applied to scalar fields propagating in a static, spherically 
symmetric, curved background. The radial part of the KG 
equation has to be rewritten in a more convenient in a systematic scheme~: the tortoise 
coordinate is used and the unknown function is redefined 
independently of the metric. The propagation of fields 
is then described by a $1-$dimensional Schr\"odinger equation whose resolution can be performed
with
a semi-classical ansatz. However, the classical problem underlying the quantum one is based on a 
slightly-modified-from-relativistic-case Hamilton-Jacobi equation where what is usually 
interpreted as the potential is here seen as the square of the potential. The 
semi-classical method was then used to study the behaviour of scalar fields in AdS and 
Schwarzschild 
backgrounds and it was shown that the energy levels in an AdS universe and the greybody factors 
entering the Hawking radiation law are both recovered at the semi-classical order with a very 
good accuracy. It should also be 
emphasized that the WKB estimation of the greybody factors provides a powerful tool for 
theoretical investigations, {\it{i.e}} for testing the influence of a given space-time curvature 
on the emission/absorption cross-sections. Finally, we have performed a semi-classical study of 
scalar fields 
propagating in SAdS space-times. The potential seen by those fields imposes resonances 
in the density of states, whose positions in energy and bandwidths can be evaluated using 
the Bohr-Sommerfeld quantization rule for the former quantity and the WKB tunnel probability for 
the latter one.\\

In principle, all the above results could be improved~: first, 
the Taylor expansion of the ansatz used in the KG equation could be preformed up to 
higher orders. Then, as it has already been mentioned 
by Langer \cite{langer}, considering the radial part of the Schr\"odinger equation as a one 
dimensional equation remains a naive approach as the centrifugal potential exhibits 
a singular behaviour when $r\to0$. As discussed in \cite{brack}, the phase of the WKB wave 
function for radial motion does not match the one of the exact wave function in regions where 
the potential can be neglected when compared to the centrifugal 
one. Nevertheless, the matching between both solutions can still 
be performed by replacing $\ell(\ell+1)$ by $(\ell+1/2)^2$ but, in any case, a precise study of 
the matching 
between the semi-classical wave functions and the exact ones, in regions where such exact 
solutions can be found, have to be carried out before applying the {\it{Langer trick}}.

Finally, the semi-classical methods provide a powerful tool to 
study the dynamics of scalar fields in curved space-times as, in most cases, exact solutions cannot 
be derived. As an example, the highly non-trivial problem of the Hawking evaporation process for 
SAdS black holes can be greatly 
simplified in this way and is currently the only available treatment. Another field of application 
is related to
primordial cosmology where the WKB approach, first used by Martin \& Schwarz \cite{martin}, has 
been extended to the second order by the authors of Ref. \cite{casadio} so as to derive the 
primordial fluctuations power spectrum. However, it has been recently proven by the authors of 
Ref. \cite{winitzki} that the accuracy of the WKB approximation is intrinsically limited
because it involves divergent series. This result, demonstrated in the case of cosmological 
particle production during the inflationary era, enlightens the cautious needed when dealing with 
WKB techniques.

\vspace{0.3cm}

\begin{flushleft}
	{\bf{Acknowledgements:}} 
\end{flushleft}

	J.G. would like to thank P. Kanti for solving the KG equation in AdS background as 
	well as L. Derome and K. Protassov for very helpful discussions.

\appendix
	\section{Appendix \ref{app-c}: Matching procedure}
	\label{app-c}

	The matching procedure involves the  equation
$
	\frac{d^2\psi}{d\xi^2}+p^2(\xi)\psi(\xi)=0
$
with two turning point at $\xi_1$ and $\xi_2$. The procedure consists in restricting the 
momentum at the linear order $p^2(\xi)\simeq-\lambda(\xi-\xi_1)$, with $\lambda>0$, and to match 
the new solutions with the WKB ones on the left and on the right of the turning point 
\cite{martin,langer,brack}. With this {\it{linearized}} momentum, the solution of the previous 
equation are expressed in terms of Airy functions $\left({\mathbf{Ai}},{\mathbf{Bi}}\right)$~:
$
	\psi(\xi)=B_1{\mathbf{Ai}}\left(\lambda^{\frac{1}{3}}(\xi-\xi_1)\right)+B_2{\mathbf{Bi}}\left(\lambda^{\frac{1}{3}}(\xi-\xi_1)\right).
$
As $\xi$ is close to $\xi_1$, this solution can be expanded in terms of exponential:
\begin{equation}
	\psi(\xi)=\left\{\begin{array}{ll}
			\frac{\lambda^{-\frac{1}{12}}}{2\sqrt{\pi}}\left|\xi-\xi_1\right|^{-\frac{1}{4}}\left[(B_2-iB_1)\exp{\left({i\frac{2}{3}\sqrt{\lambda}\left|\xi-\xi_1\right|^{\frac{3}{2}}+i\frac{\pi}{4}}\right)}\right. & \\
			\left.+(B_2+iB_1)\exp{\left(-i\frac{2}{3}\sqrt{\lambda}\left|\xi-\xi_1\right|^{\frac{3}{2}}-i\frac{\pi}{4}\right)}\right] & ~~\mathrm{in~region}~\mathcal{A} \\
			\\
			\frac{\lambda^{-\frac{1}{12}}}{\sqrt{\pi}}\left|\xi-\xi_1\right|^{-\frac{1}{4}}\left[\frac{B_1}{2}\exp{\left(-\frac{2}{3}\sqrt{\lambda}\left|\xi-\xi_1\right|^{\frac{3}{2}}\right)}\right. & \\
			\left.+B_2\exp{\left(\frac{2}{3}\sqrt{\lambda}\left|\xi-\xi_1\right|^{\frac{3}{2}}\right)}\right] & ~~\mathrm{in~region}~\mathcal{B}. \\
		    \end{array}\right.
\end{equation}
To perform the matching between Airy functions solutions and the WKB solutions, we calculate the 
WKB solutions in the two region using the linear approximation for the momentum~:
\begin{equation}
	\psi(\xi)=\left\{\begin{array}{ll}
			\lambda^{-\frac{1}{4}}\left|\xi-\xi_1\right|^{-\frac{1}{4}}\left[A_{out}\exp{\left({-i\frac{2}{3}\sqrt{\lambda}\left|\xi-\xi_1\right|^{\frac{3}{2}}}\right)}\right. & \\
			\left.+A_{in}\exp{\left({i\frac{2}{3}\sqrt{\lambda}\left|\xi-\xi_1\right|^{\frac{3}{2}}}\right)}\right] & ~~\mathrm{in~region}~\mathcal{A} \\
			\\
			\lambda^{-\frac{1}{4}}\left|\xi-\xi_1\right|^{-\frac{1}{4}}A_{eva}\exp{\left(-\frac{2}{3}\sqrt{\lambda}\left|\xi-\xi_1\right|^{\frac{3}{2}}\right)} & ~~\mathrm{in~region}~\mathcal{B}. \\
		    \end{array}\right.
\end{equation}
Matching $B_{1/2}$ to $A_{out/in}$ and to $A_{eva/div}$, it is possible to match the two solutions
in regions ${\mathcal{A}}$ and ${\mathcal{B}}$~: $A_{eva}=e^{-i\frac{\pi}{4}}A_{out}$ and 
$A_{in}=e^{-i\frac{\pi}{2}}A_{out}$.

	The same procedure is applied at the second turning point $\xi_2$ just by introducing 
	$\xi'=-\xi$, at this point the WKB solutions take the form~:
	\begin{equation}
	\psi(\xi')=\left\{\begin{array}{ll}
			\lambda^{-\frac{1}{4}}\left|\xi'-\xi'_2\right|^{-\frac{1}{4}}A_{eva}e^{-\tau}\exp{\left(\frac{2}{3}\sqrt{\lambda}\left|\xi'-\xi'_2\right|^{\frac{3}{2}}\right)}& ~~\mathrm{in~region}~\mathcal{B} \\
			 & \\
			\lambda^{-\frac{1}{4}}\left|\xi'-\xi'_2\right|^{-\frac{1}{4}}B_{out}\exp{\left({i\frac{2}{3}\sqrt{\lambda}\left|\xi'-\xi'_2\right|^{\frac{3}{2}}}\right)} & ~~\mathrm{in~region}~\mathcal{C} \\
		    \end{array}\right.
\end{equation}
with $\tau=\displaystyle\int_{r^\star_1}^{r^\star_2}{p'(\xi)d\xi}$ and gives: $B_{out}=e^{i\frac{\pi}{4}}e^{-\tau}A_{eva}$.  From this equation and the previous one, we finally obtain the amplitude of the outgoing waves at $+\infty$ in function of the amplitude of the outgoing waves at $-\infty$:
$
	B_{out}=e^{-\tau}A_{out}.
$

	\section{Appendix \ref{app-a}: Resolution of the AdS-KG equation}
	\label{app-a}

	To solve  \cite{private} the differential Eq. (\ref{eq-ads}), a change of variable as 
	well as a change of the unknown function is performed~:
$
		z=\cos{(2\rho)}~;~
		R(\rho)=\sin^\ell{(\rho)}\cos^\beta{(\rho)}P(\rho).
	$
Under this change, the differential equation is of the form \cite{stegun}
$
	(1-z^2)\frac{d^2P}{dz^2}+(b-a-(a+b+2)z)\frac{dP}{dz}+n(n+a+b+1)P(z)=0
	\label{eq-ads2}
$
where the parameters $(\beta,a,b,n)$ must satisfy the four constraints~:
\begin{equation}
	\begin{array}{l}
		a+b+2=\ell+1+\beta~;~
		b-a=\beta-2-\ell~;~
		\beta(\beta-3)=0~;~
		n(n+a+b+1)=\frac{1}{4}\left(\nu^2-(\ell+\beta)^2\right).
	\end{array}
	\label{constrain}
\end{equation}

The solution are the Jacobi's polynomials $P_n^{(a,b)}(z)$, imposing that 
$n$ is an integer, $a>-1$ and $b>-1$ \cite{stegun}. Once the four constraints for the parameters 
$(\beta,a,b,n)$ solved, the solution of the radial part of the KG equation is expressed in 
function of the Jacobi's polynomials~:
$
	R(\rho)=\sin^\ell{(\rho)}\cos^3{(\rho)}P_{(\nu-\ell-3)/2}^{(\ell+1/2,3/2)}\left(\cos{(2\rho)}\right)
$
where we choose $\beta=3$ in order to respect the $b>-1$ constraint.

	\section{Appendix \ref{app-b}: Determination of the turning points}
	\label{app-b}

	In the $\ell\neq0$ case, the determination of $r^\star_{\pm}$ is more complicated. 
	Introducing $y=\tan{({r^\star}/{R})}$, the solutions of the equation $\omega=V(r)$ are 
	given by the algebraic equation
$
	2y^4+\left(2+\ell(\ell+1)-\omega^2R^2\right)y^2+\ell(\ell+1)=0.
$
In order to have two different, positive, real roots, three constraints have to be satisfied:
\begin{equation}
	\begin{array}{l}
		2+\ell(\ell+1)-\omega^2R^2\leq0 \\
		\left(2+\ell(\ell+1)-\omega^2R^2\right)^2-8\ell(\ell+1)\geq0 \\
		8\ell(\ell+1)\geq0
	\end{array}
\end{equation}
where the third one is always satisfied. Solving the first constraint, the zero-point energy 
feature of the energy spectrum is recovered $\omega\geq\frac{1}{R}\sqrt{2+\ell(\ell+1)}$. This result is in good agreement with the exact energy spectrum as $\sqrt{2+\ell(\ell+1)}<\ell+3$. Under such constraints, only two positive, real roots exist which are the two turning points:
\begin{equation}
	y_{\pm}=\frac{1}{2}\sqrt{\omega^2R^2-2-\ell(\ell+1)\pm\sqrt{\left(2+\ell(\ell+1)-\omega^2R^2\right)^2-8\ell(\ell+1)}}.
\end{equation}

\end{document}